\documentstyle[aps,prb,epsfig,amssymb,floats]{revtex}

\begin{document}
\draft

\twocolumn[\hsize\textwidth\columnwidth\hsize\csname
@twocolumnfalse\endcsname

\title{Phonons and Lattice Dielectric Properties of Zirconia}
\author{Xinyuan Zhao and David Vanderbilt}
\address{Department of Physics and Astronomy, Rutgers University,
	Piscataway, NJ 08854-8019}
\date{August 28, 2001}
\maketitle

\begin{abstract}
We have performed a first-principles study of the structural and
vibrational properties of the three low-pressure (cubic,
tetragonal, and especially monoclinic) phases of ZrO$_2$, with
special attention to the computation of the zone-center phonon
modes and related dielectric properties.  The calculations have
been carried out within the local-density approximation using
ultrasoft pseudopotentials and a plane-wave basis.  The fully
relaxed structural parameters are found to be in excellent
agreement with experimental data and with previous theoretical
work. The total-energy calculations correctly reproduce the
energetics of the ZrO$_{2}$ phases, and the calculated zone-center
phonon frequencies yield good agreement with the infrared and
Raman experimental frequencies in the monoclinic phase.
The Born effective charge tensors are computed and, together with
the mode eigenvectors, used to decompose the lattice dielectric
susceptibility tensor into contributions arising from individual
infrared-active phonon modes.  This work has been partially
motivated by the potential for ZrO$_{2}$ to replace SiO$_{2}$
as the gate-dielectric material in modern integrated-circuit
technology. 
\end{abstract}


\vskip2pc]

\columnseprule 0pt

\narrowtext

\section{Introduction}
\label{sec:intro}

ZrO$_{2}$, or zirconia, has a wide range of materials
applications because of its high strength and stability at high
temperature.  A prospective application of particular
current interest is its possible use to replace SiO$_{2}$ as the
gate-dielectric material in metal-oxide-semiconductor (MOS)
devices. 

The use of SiO$_2$ as the gate dielectric, and in particular the
quality of the Si/SiO$_{2}$ interface, have been a foundation
of modern integrated-circuit technology since its invention decades
ago.  Driven by the
seemingly endless pressure for higher operation speed, smaller
physical dimensions, and lower driving voltage, 
the gate dielectric thickness in integrated circuits has been rapidly
reduced from the order of 1$-$2$\,\mu$m in the early 1960s to the current
value of about 2$-$3\,nm. If SiO$_2$ is not replaced by another
material, this would require the gate dielectric thickness to be
reduced to less than 1\,nm in the coming 
decade.\cite{garfunkel}  Such a reduction in gate oxide
thickness, however, would impose several severe problems on the
current Si/SiO$_{2}$ semiconductor technology, including a high
level of direct tunneling current, a large degree of dopant
(boron) diffusion in the gate oxide, and reliability problems
associated with nonuniformity of the very thin film.  It has been
demonstrated that the direct tunneling current grows
exponentially as the thickness of the gate dielectric film
decreases.\cite{lo97,heiser97}  For films thinner than 2\,nm,
the tunneling current could become as large as 1\,A/cm$^{2}$,
which would require a level of power dissipation that would be
intolerable for most digital device applications.\cite{momose96}
These fundamental problems are largely attributable to the
inherently low dielectric constant of silicon dioxide ($\epsilon
\simeq 3.5$), quite small in comparison with many other oxide
dielectrics.

Several approaches have been proposed for overcoming these
fundamental challenges associated with the use of SiO$_{2}$ films.
In particular, much recent effort has been focused on metal oxides
having a larger dielectric constant than that of SiO$_{2}$, since
these might be used to provide physically thicker dielectric
films that are equivalent to much thinner SiO$_2$ ones in terms
of their capacitance, but exhibiting a greatly reduced leakage
current.  Some of the proposed candidates include
Ta$_{2}$O$_{5}$,\cite{alers98,grahn98} TiO$_{2}$, ZrO$_{2}$,
Y$_{2}$O$_{3}$, Al$_{2}$O$_{3}$, and hafnium and zirconium
silicate systems (Hf$_{1-x}$Si$_{x}$O$_{2}$ and
Zr$_{1-x}$Si$_{x}$O$_{2}$).\cite{wilk00}  Among these candidates,  
ZrO$_2$ is a promising one because of its good dielectric
properties ($\epsilon \sim 20$) and thermodynamic
stability in contact with the Si substrate.

Zirconia is known to have three low-pressure structural phases.
The system passes from the monoclinic ground state to a tetragonal
phase, and then eventually to a cubic phase, with increasing
temperature.  The monoclinic phase (space group $C^{5}_{2h}$ or
$P2_{1}/c$) is thermodynamically stable below 1400\,K. Around 1400\,K
a transition occurs to the tetragonal structure (space group
$D^{15}_{4h}$ or $P4_{2}/nmc$), which is a slightly distorted
version of the cubic structure and is stable up to 2570\,K. Finally,
the cubic phase (space group $O_{h}^{5}$ or $Fm3m$) is thermodynamically
stable between 2570\,K and the melting temperature at 2980\,K. This
information is summarized in Table~\ref{table:phases}, which also
shows the coordination number of the Zr and O atoms for each of
the three phases. In the monoclinic phase there are two
nonequivalent oxygen sites with coordination numbers of 3 (O$_{1}$)  
and 4 (O$_{2}$), while all the Zr atoms are equivalent and have a
coordination of 7.

Our purpose is to investigate the lattice contributions to the
dielectric properties of these three ZrO$_{2}$ phases, especially the
monoclinic phase.  Because previous experimental and theoretical work
indicates that the electronic contribution to the dielectric
constant is rather small ($\epsilon_\infty\simeq 5$) and is neither
strongly anisotropic nor strongly dependent on structural phase,
\cite{gonze,feinberg,french,nassau,rignanese2} and 
because $\epsilon_\infty$ is best calculated by specialized
linear-response techniques, we have not calculated it here.
Instead, we focus on the lattice contributions to the dielectric
response because, as we shall see, these are much larger, more
anisotropic, and more sensitive to the lattice structure.

\begin{table}
\caption{The three low-pressure phases of ZrO$_{2}$.
The last three columns give the coordination numbers of the
Zr and O atoms. (Atoms O$_1$ and O$_2$ are equivalent in
the cubic and tetragonal, but not in the monoclinic, structures.)}
\begin{tabular}{lccccc}

 & & & \multicolumn{3}{c}{Coordination} \\
\mbox{Phase} & \mbox{Space group} & \mbox{$T$\,(K)} &
               \mbox{Zr} & \mbox{O$_1$} & \mbox{O$_2$} \\  
\tableline
\mbox{Cubic}      & $Fm3m$  & \mbox{2570 - 2980} & 8 & 4 & 4 \\
\mbox{Tetragonal} & $P4_{2}/nmc$ & \mbox{1400 - 2570} & 8 & 4 & 4 \\
\mbox{Monoclinic} & $P2_{1}/c$   & $<$1400            & 7 & 3 & 4   

\end{tabular}
\label{table:phases}
\end{table}

In order to achieve this, the Born effective charge  
tensors and the force-constant matrices are calculated for the three
ZrO$_{2}$ phases using density-functional theory.
We first check that our relaxed structural parameters and energy
differences between phases are consistent with previous theoretical
\cite{stefanovich94,wilson96,dewhurst98,louie98,jomard99,%
stapper99,fabris00,jansen}
and experimental work.\cite{howard,aldebert}
The Born effective charge tensors are then computed from finite
differences of polarizations as various sublattice displacements
are imposed, with the polarizations computed using the
Berry-phase method.\cite{modern-pol} The force constants are
obtained in a similar way from finite differences of forces.
Reasonable agreement is found between the calculated frequencies and the 
measured spectra for both IR-active and Raman-active modes,
\cite{feinberg,asher75,hirata,carlone,zhang99} although
possible reassignments are proposed for certain modes based on the
results of our calculations.   Finally, our theoretical information
is combined to predict the lattice contributions to the bulk
dielectric tensor.   We thus clarify the dependence of the
dielectric response on crystal phase, orientation, and lattice
dynamical properties. In particular, we find that the lattice
dielectric tensors in the tetragonal and monoclinic phases are strongly
anisotropic.  We also find that the monoclinic phase has
the smallest orientationally-averaged dielectric constant of the three
phases, owing to the fact that the mode effective charges associated
with the lowest-frequency modes are rather weak.

The paper is organized as follows. In Sec.~\ref{sec:details} we
briefly describe the technical aspects of our first-principles
calculations. Sec.~\ref{sec:results} presents the results, including
the structural relaxations, the Born effective charge tensors, the
phonon normal modes, and the lattice contributions to the dielectric
tensors. Sec.~\ref{sec:conclusion} concludes the paper.

\section{Details of First-Principles Calculations}
\label{sec:details}

The calculations are carried out within a plane-wave pseudopotential
implementation of density-functional theory (DFT)
in the local-density approximation (LDA) using
Ceperley-Alder exchange-correlation.\cite{alder,explan-gga} 
The use of Vanderbilt ultrasoft pseudopotentials \cite{ultrasoft} 
allows a highly accurate calculation to be achieved with a low energy
cutoff, which is chosen to be 25\,Ry in this work. 
The $4s$ and $4p$ semicore shells
are included in the valence for Zr, and the $2s$ and $2p$ shells are 
included in the valence for O.  A conjugate-gradient algorithm is
used to compute the total energies and forces.
For each of the three ZrO$_{2}$ phases, a unit cell containing 12 atoms
(4 Zr and 8 O atoms) is used in our calculations.  Although we thus
use an unnecessarily large cell for the cubic and tetragonal phases, 
this approach has the advantage that the three zirconia phases can be  
studied in a completely parallel fashion.

A 4$\times$4$\times$4 Monkhorst-Pack \cite{monk-pack} k-point mesh is found to
provide sufficient precision in the calculations of total energies
and forces.  In order to calculate Born effective charges and
force-constant matrices, each atomic sublattice in turn is
displaced in each Cartesian direction by $\pm$0.2\% in lattice
units, and the Berry-phase polarization \cite{modern-pol} and
Hellmann-Feynman forces are computed. To be specific,  
a 4$\times$4$\times$20 k-point sampling over the Brillouin zone was
used in the Berry-phase polarization calculations, and we have confirmed that 
good convergence was achieved for the three ZrO$_2$ phases 
with such k-point sampling. The Born effective charge  
tensors and force-constant matrices are then constructed by finite
differences from the results of these calculations.

\begin{figure}
\begin{center}
   \epsfig{file=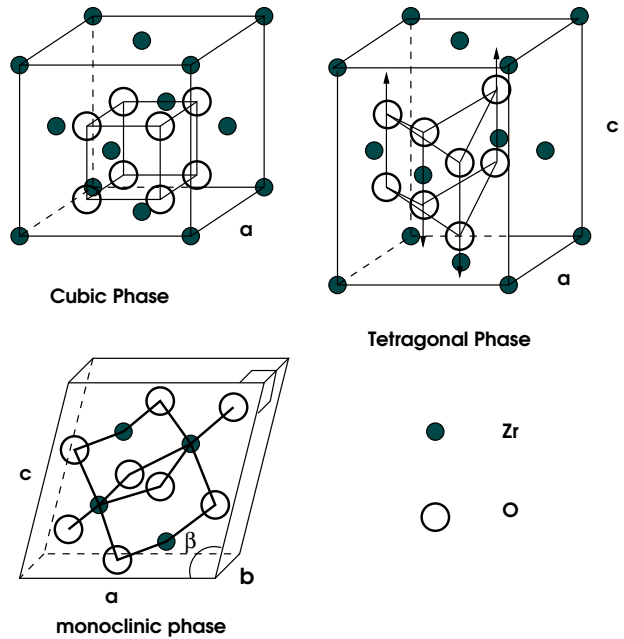,width=3.2in}
\end{center}
\caption{Structures of the three ZrO$_{2}$ phases. The
Zr$-$O bonds are only shown in the monoclinic structure. For the
tetragonal phase, the arrows indicate the distortion of oxygen pairs
relative to the cubic structure.}
\label{fig:struct}
\end{figure}

\section{Results}
\label{sec:results}

\subsection{Atomic Structures of ZrO$_{2}$ Phases}
\label{sec:struct}

The three crystal structures of ZrO$_2$ are shown in Fig.~\ref{fig:struct}.
Cubic zirconia takes the fluorite (CaF$_{2}$) structure,
in which the Zr atoms are in a face-centered cubic
structure and the oxygen atoms occupy the tetrahedral interstitial
sites associated with this fcc lattice. The structure of tetragonal
zirconia can be regarded as a distortion of the
cubic structure obtained by displacing alternating pairs of oxygen atoms
up and down by an amount $\Delta z$ along the $z$ direction, as shown
in the figure.  This doubles the primitive cell from three to six atoms
and is accompanied by a tetragonal strain.  The structure can be
specified by the two lattice parameters $a$ and $c$ and a
dimensionless ratio $d_{z} = \Delta z / c$.  Cubic zirconia can be
considered as a special case of the tetragonal structure with
$d_{z}=0$ and $c/a = 1$ (if the primitive cell is used for tetragonal
phase, $c/a = \sqrt 2$).

Monoclinic zirconia has a lower symmetry and a more complex geometric 
structure with a 12-atom primitive cell. The lattice parameters are
$a$, $b$, $c$, and $\beta$ (the 
non-orthogonal angle between $\bf a$ and $\bf c$)
as shown in Fig.~\ref{fig:struct}. The
atomic coordinates in Wyckoff (lattice-vector) notation
are $\pm (x,y,z)$ and $\pm (-x,y+1/2,1/2-z)$, with
parameters $x$, $y$ and $z$ specified for each of three kinds of
atoms: Zr, O$_1$, and O$_2$.  
Note that there are two non-equivalent oxygen sites: atoms of type
O$_{1}$ are 3-fold coordinated, while O$_2$ are 4-fold coordinated.
All Zr atoms are equivalent and are 7-fold coordinated.
Thus, four lattice-vector parameters and nine internal parameters
are needed to specify the structure fully.

\begin{table}
\caption{Structural parameters obtained for three ZrO$_{2}$ phases
from present theory, compared with previous pseudopotential (PP)
and linear augmented plane-wave (FLAPW) calculations and with
experiment.  Lattice parameters $a$, $b$, $c$ and volume per
formula unit $V$ are in atomic units; monoclinic angle $\beta$ is
in degrees; and internal coordinates $d_z$, $x$, $y$ and $z$ are
dimensionless.}
\begin{tabular}{cdddd}
& \mbox{This work} & \mbox{PP}\tablenotemark[1]
                     & \mbox{FLAPW}\tablenotemark[2]
                     & \mbox{Expt.}\tablenotemark[3] \\
\tableline
\multicolumn{5}{l}{Cubic} \\
\mbox{$V$} & 215.612 & 215.31 & 217.79 & 222.48 \\
\mbox{a}      & 9.5187  & 9.514  & 9.551  & 9.619 \\
\tableline
\multicolumn{5}{l}{Tetragonal} \\
\mbox{$V$} & 217.698 & 218.69 & 218.77 & 222.96 \\
\mbox{a}      & 9.5051   & 9.523  & 9.541  & 9.543 \\
\mbox{c}      & 9.6383   & 9.646  & 9.613  & 9.793 \\
$d_{z}$ & 0.0418  & 0.0423 & 0.029  & 0.0574 \\
\tableline
\multicolumn{5}{l}{Monoclinic} \\
\mbox{$V$} & 231.822 & 230.51 &        & 237.71 \\
\mbox{a}    & 9.6532 & 9.611  & & 9.734 \\
\mbox{b}    & 9.7690 & 9.841  & & 9.849 \\
\mbox{c}    & 9.9621 & 9.876  & & 10.048 \\
$\beta$     & 99.21  & 99.21  & & 99.23 \\
$x_{\rm Zr}$    & 0.2769 & 0.2779 & & 0.2754 \\
$y_{\rm Zr}$    & 0.0422 & 0.0418 & & 0.0395 \\
$z_{\rm Zr}$    & 0.2097 & 0.2099 & & 0.2083 \\
$x_{\rm O_{1}}$ & 0.0689 & 0.0766 & & 0.0700 \\
$y_{\rm O_{1}}$ & 0.3333 & 0.3488 & & 0.3317 \\
$z_{\rm O_{1}}$ & 0.3445 & 0.3311 & & 0.3447 \\
$x_{\rm O_{2}}$ & 0.4495 & 0.4471 & & 0.4496 \\
$y_{\rm O_{2}}$ & 0.7573 & 0.7588 & & 0.7569 \\
$z_{\rm O_{2}}$ & 0.4798 & 0.4830 & & 0.4792 \\
\end{tabular}
\tablenotetext[1] {Ref.~\onlinecite{louie98}.}
\tablenotetext[2] {Ref.~\onlinecite{jansen}.}
\tablenotetext[3] {Ref.~\onlinecite{stefanovich94}.}
\label{table:parameters}
\end{table}

Tabulated in Table~\ref{table:parameters} are the relaxed structural
parameters for the three phases of ZrO$_{2}$ as computed within our
energy minimization procedure, as well as results of previous
theoretical and experimental work for comparison.  The experimental
parameters given in the last column were used as the starting point
for our DFT--LDA structural relaxations.
It can readily be seen that there is excellent agreement between our
results and previous theory and experiment.
The volumes are all slightly underestimated, by 2-3\%, as is typical of LDA
calculations. The largest discrepancy is for $d_z=\Delta z/c$, the internal
coordinate in the tetragonal phase; our value is
$\sim$30\% smaller than the experimental value, but it is
very closed to the results of the previous
pseudopotential calculation.  (The discrepancy with experiment should
not be taken too seriously, in view of the fact that the theory is
a zero-temperature one.)
The very close (usually $<$\,1\%) agreement with the previous
pseudopotential results of Ref.~\onlinecite{louie98} provides
a good confirmation of the reliability of our calculations.

\begin{table}
\caption{O--Zr bond lengths and Zr--O--Zr bond angles in monoclinic
zirconia (in \AA\ and degrees respectively).  Values in parentheses
are from {Ref.~\protect\onlinecite{trueblood}} for comparison.}
\begin{tabular}{cdddd}
\multicolumn{4}{l}{O$_{1}$-Zr bond lengths and angles} \\ \tableline
$d_1$ & 2.035 & (2.051) & $\theta_{12}$ & 138.6 \\
$d_2$ & 2.051 & (2.057) & $\theta_{13}$ & 106.3 \\
$d_3$ & 2.144 & (2.151) & $\theta_{23}$ & 105.0 \\ \tableline
\multicolumn{4}{l}{O$_{2}$-Zr bond lengths and angles} \\ \tableline
$d_1$ & 2.138 & (2.163) & $\theta_{12}$ & 108.6 \\
$d_2$ & 2.229 & (2.220) & $\theta_{13}$ & 106.0 \\
$d_3$ & 2.153 & (2.189) & $\theta_{14}$ & 133.0 \\
$d_4$ & 2.233 & (2.285) & $\theta_{23}$ & 102.0 \\
      &       &         & $\theta_{24}$ & 100.6 \\
      &       &         & $\theta_{34}$ & 103.6
\end{tabular}
\label{table:bond}
\end{table}

\begin{figure}
\begin{center}
   \epsfig{file=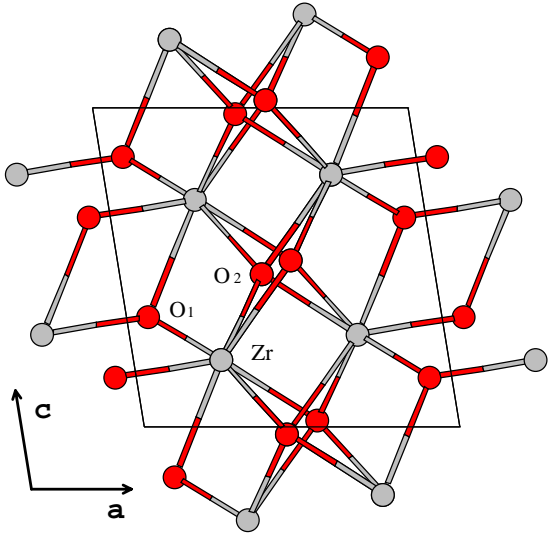, width=2.9in}
\end{center}
\caption{Relaxed lattice structure of monoclinic
ZrO$_{2}$; unit cell is outlined.  Light and dark 
circles stand for the Zr and O atoms, respectively. A 3-fold
coordinated oxygen atom (O$_{1}$) is bonded to the nearest neighboring
Zr atoms in an almost planar configuration, while a 4-fold oxygen
(O$_{2}$) forms a distorted tetrahedron with the Zr neighbors.}
\label{fig:mono}
\end{figure}

Fig.~\ref{fig:mono} illustrates the relaxed monoclinic structure,
and Table~\ref{table:bond} lists the calculated bond lengths and bond
angles for the O--Zr bonds.  Bond lengths taken from Ref.~\onlinecite{trueblood}
are also listed for comparison.  As can be seen in the figure, a
three-fold coordinated oxygen atom (O$_{1}$) is bonded to the three
nearest-neighbor Zr atoms in an almost planar configuration,
as can be verified by noting that the sum of the three bond angles
is about 350$^{\circ}$.  A second four-fold oxygen atom (O$_{2}$) forms a
distorted tetrahedron with its four nearest Zr neighbors, the degree
of distortion being evident from the lengths and angles in the table.
The presence of these two distinct oxygen atoms with utterly different
environments suggests that their contributions to the dielectric
properties of the material may be quite different.  We shall see how
this is manifest as a difference of the Born effective charge tensors
for O$_1$ and O$_2$ in the next subsection.

Our total-energy calculations have correctly reproduced the energetics of the 
three ZrO$_2$ phases. The differences of total energies per formula unit
for the monoclinic and tetragonal phases, relative to the cubic phase,  
are 0.044\,eV and 0.089\,eV respectively, to be compared with 0.045\,eV 
and 0.102\,eV from previous calculation,\cite{louie98} and 0.057\,eV and  
0.120\,eV from one experiment.\cite{ackermann}

\subsection{Born Effective Charge Tensors}
\label{sec:born}

The Born effective charge tensor quantifies the macroscopic electric
response of a crystal to internal displacements of its atoms.
We begin with a calculation of the bulk
polarization {\bf P}, using the Berry-phase polarization method to
compute the electronic contribution, as formulated in
Ref.~\onlinecite{modern-pol}.
${\bf Z}_{i}^{*}$, the Born effective charge tensor
for the $i$-th atom in the unit cell, is defined via
\begin{equation}
\label{eq:born_eff}
\Delta {\bf P} = \frac{e}{V} \sum_{i=1}^{N} {\bf Z}_{i}^{*} \cdot \Delta
{\bf u}_{i}
\end{equation}
where $V$ is the volume of the unit cell, $\Delta {\bf u}_{i}$ is the
displacement of the $i$-th atom in the unit cell, and $\Delta {\bf P}$
is the induced change in bulk polarization resulting from this
displacement.  Using Eq.~(\ref{eq:born_eff}),
$\bf{Z}^{*}$ can be computed from finite differences 
of {\bf P} under small but finite distortions.\cite{resta} 

In the Berry-phase polarization scheme, one samples the Brillouin zone
by a set of strings of k-points set up parallel to some chosen reciprocal
lattice vector, thereby computing the electronic polarization along
that direction. For cubic and tetragonal ZrO$_{2}$, this is relatively
straightforward since the reciprocal lattice vectors are all mutually
perpendicular.  For monoclinic ZrO$_{2}$, however, one has to transform
the polarization to Cartesian coordinates after first computing it in
lattice coordinates.

\begin{table}
\caption{Born effective charges for three phases of ZrO$_2$.  In the
cubic phase, the ${\bf Z}^*$ tensors are diagonal and isotropic.
In the tetragonal phase, the ${\bf Z}^*$ tensors are diagonal in an
$x'$--$y'$--$z$ frame rotated 45$^\circ$ about $\hat{z}$ from the
Cartesian frame; $Z^*_j$ ($j$=1,2,3) are $Z^*_{x'x'}$, $Z^*_{y'y'}$,
and $Z^*_{zz}$, respectively.  In the monoclinic phase, $Z^*_j$ is
the $j$'th eigenvalue of the symmetric part of the ${\bf Z}^*$ tensor.}
\begin{tabular}{lcddd}
\mbox{Phase} & \mbox{Atom} &$Z^{*}_1$ & $Z^{*}_2$ & $Z^{*}_3$ \\
\tableline
\mbox{Cubic} & \mbox{Zr} & 5.72 & 5.72 & 5.72 \\
 & \mbox{O}  & \mbox{$-$2.86} & \mbox{$-$2.86} & \mbox{$-$2.86} \\
\tableline
\mbox{Tetragonal} & \mbox{Zr} & 5.75 & 5.75 & 5.09 \\
 & \mbox{O$_{1}$} & \mbox{$-$3.53} & \mbox{$-$2.22} & \mbox{$-$2.53} \\
 & \mbox{O$_{2}$} & \mbox{$-$2.22} & \mbox{$-$3.53} & \mbox{$-$2.56} \\
\tableline
\mbox{Monoclinic} & \mbox{Zr} & 4.73 & 5.42 & 5.85  \\
 & \mbox{O$_{1}$} & \mbox{$-$4.26} & \mbox{$-$2.64} & \mbox{$-$1.19}  \\
 & \mbox{O$_{2}$} & \mbox{$-$3.20} & \mbox{$-$2.52} & \mbox{$-$2.26} 
\end{tabular}
\label{table:born_ct}
\end{table}

Our results for the dynamical effective charges of the three phases
are presented in Table \ref{table:born_ct}.
In the cubic phase, symmetry requires that the Born effective charge
tensor should be isotropic ($Z^*_{ij}=Z^*\,\delta_{ij}$) on each atom,
and that $Z^*$(O$_1$)=$Z^*$(O$_2$); the neutrality sum rule requires
that $Z^*$(Zr)=$-2 \, Z^*$(O).  The values given in Table \ref{table:born_ct}
can be seen to be in excellent agreement with the corresponding values of
$Z^*$(Zr)=5.75 and $Z^*$(O)=$-$2.86 reported in Ref.~\onlinecite{gonze}.

In the tetragonal phase, ${\bf Z}^*$(Zr) is diagonal in the Cartesian
frame with $Z^*_{xx}$=$Z^*_{yy}$$\ne$$Z^*_{zz}$. The diagonal elements
of ${\bf Z}^*$(O) have the same form, but the shifting of oxygen atom
pairs creates two different configurations for oxygen atoms (denoted
O$_{1}$ and O$_{2}$) and introduces off-diagonal $xy$ elements.  Specifically,
$Z^*_{xy}$(O$_1$)=$Z^*_{yx}$(O$_1$)=$-Z^*_{xy}$(O$_2$)=$-Z^*_{yx}$(O$_2$).
Thus, it is more natural to refer to a reference frame that has been
rotated 45$^\circ$ about the $\hat{z}$ axis; in this frame the ${\bf Z}^*$(O)
become diagonal.  This symmetry analysis is confirmed in our
calculations, as can be seen from Table~\ref{table:born_ct}. 
We have recently become aware of the independent work of
Ref.~\onlinecite{rignanese2}, which also reports values for the ${\bf Z}^*$
tensors in the tetragonal phase of ZrO$_{2}$.  These authors
find $Z^*_{xx}$=5.74 and $Z^*_{zz}$=5.15 for Zr, and
$Z^*_{x'x'}$=$-$3.52, $Z^*_{y'y'}$=$-$2.49 and $Z^*_{zz}$=$-$2.57
for oxygens. Evidently there is again very good agreement between our
results and those of previous theory.

In the monoclinic phase, the Born effective charge tensors are more
complicated because of the complexity of the lattice structure. The
two oxygen sites are now non-equivalent, and the crystal structure should
be regarded as composed of three kinds of atoms, namely, Zr, O$_{1}$,
and O$_{2}$.  Each kind of atom appears four times in the unit cell,
once at a ``representative'' Wyckoff position $(x,y,z)$, and then also at
partner positions $(-x, -y, -z)$, $(-x, 0.5+y, 0.5-z)$ and $(x,
0.5-y, 0.5+z)$ given by action of the space-group operations
$E$, $I$, \{$C_{2}^{y} \,|\, 0, 0.5, 0.5$\} and \{$M_{y} \,|\, 0,
0.5, 0.5$\}.  Thus, all three kinds of atoms have equally low
symmetry, and their resulting ${\bf Z}^*$ tensors are neither
diagonal nor symmetric.  Specifically, for these representative
atoms we find

\[   Z^{*}({\rm Zr}) =  \left( \begin{array}{rrr}
        5.471 & -0.432 & 0.180 \\
        -0.155 & 5.608 & 0.152 \\
        0.197 & 0.376 & 4.952
          \end{array} \right) \]

\[    Z^{*}({\rm O_{1}}) =  \left( \begin{array}{rrr}
        -3.019 & 1.172 & -0.199 \\
        1.449 & -2.755 & -0.695 \\
        -0.191 & -0.684 & -2.321 \\
          \end{array} \right)  \]

\[    Z^{*}({\rm O_{2}}) =  \left( \begin{array}{rrr}
        -2.461 & 0.171 & 0.018 \\
        0.238 & -2.850 & 0.372 \\
        -0.019 & 0.413 & -2.657 \\
          \end{array} \right)  \]

\noindent
We have confirmed that our computed effective-charge tensors for
the other atoms obey the relations expected by symmetry, namely,
that the ${\bf Z}^*$ tensors should be identical for partners at
$(-x, -y, -z)$, and that the off-diagonal $xy$, $yx$, $yz$, and
$zy$ matrix elements should change sign for the partners at $(-x,
0.5+y, 0.5-z)$ and $(x, 0.5-y, 0.5+z)$.  In Table~\ref{table:born_ct}
we report the eigenvalues of the symmetric part of the effective-charge
tensors.

It is obvious from Table~\ref{table:born_ct} that the Z$^{*}$ values
are quite different from the nominal ionic valences
($+$4 for Zr and $-$2 for O). Except for the value of
$-$1.19, all other magnitudes are greater than their nominal
valences. The anomalously large $Z^{*}$ values indicate that there
is a strong dynamic charge transfer along the Zr$-$O bond as the
bond length varies, indicating a mixed ionic-covalent nature of the Zr$-$O
bond. Such an anomaly reflects the relatively delocalized structure of
the electronic charge distributions, and is quite common in other
weakly ionic oxides such as the ferroelectric perovskites.\cite{zhong}

\begin{figure}
\begin{center}
   \epsfig{file=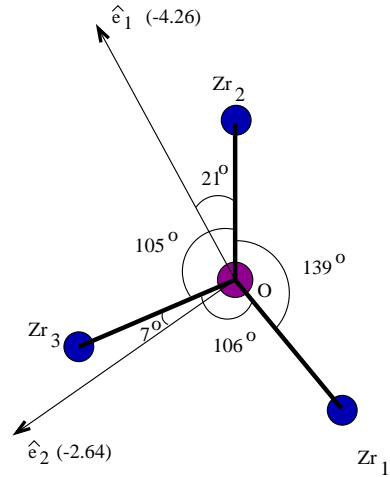, width=2.0in}
\end{center}
\caption{Environment of three-fold coordinated
O$_1$ atom in the monoclinic phase.
The three Zr$-$O bonds lie approximately in a plane.
$\hat{e}_{1}$ and $\hat{e}_{2}$ are the two principal axes
associated with the eigenvalues $-$4.26 and $-$2.64 of the
symmetric part of the $Z^*$ tensor, respectively.}
\label{fig:3_fold}
\end{figure}

As discussed in Sec.~\ref{sec:struct}, the oxygen atom of type
O$_{1}$ is bonded to three nearest-neighbor Zr atoms in an almost
planar configuration. One might then expect that the largest
dynamical charge transfer would occur for motions of the O atom
in this plane, with a smaller magnitude of $Z^*$ for motion
perpendicular to this plane.  To check this, we computed the
eigenvectors that result from diagonalizing the symmetric part of
the Born charge tensor of the O$_{1}$ atom, corresponding to the
eigenvalues in the penultimate row of Table~\ref{table:born_ct}.
Sure enough, the principle axis $\hat{e}_3$ associated with the
eigenvalue $Z^*_3$=$-$1.19 of smallest magnitude points almost directly
normal to the plane of the neighbors (making angles of $85^{\circ}$,
$91.2^{\circ}$ and $93.9^{\circ}$ to the three O--Zr bonds).  The
other two principal axes lie essentially in the plane of the
neighbors, as shown in Fig.~\ref{fig:3_fold}.  Moreover, the principal
axis $\hat{e}_1$ connected with the eigenvalue $Z^*_1$=$-$4.26
of largest magnitude is nearly parallel to the bond to the closest
neighbor Zr$_1$.  It can also be seen that the vector
$\hat{e}_2$ connected with the intermediate eigenvalue
is very nearly aligned with the O$_1$--Zr$_3$ bond.
Not surprisingly in view of its more tetrahedral coordination,
the $Z^*$ tensor for atom O$_2$ is more isotropic, as indicated
by the smaller spread of the eigenvalues in the last line of
Table~\ref{table:born_ct}.

\subsection{Phonons}
\label{sec:phonons}

The frequencies of phonons at $\Gamma$, the center of the Brillouin zone,
are calculated for the cubic, tetragonal and monoclinic phases.
For each phase, we first calculate the force-constant matrix
\begin{equation}
\Phi^{\alpha \beta}_{ij} = - \frac{\partial F^{\alpha}_{i}}{\partial
u^{\beta}_{j}} \simeq - \frac{\Delta F^{\alpha}_{i}}{\Delta
u^{\beta}_{j}} 
\end{equation}
obtained by calculating all the Hellmann-Feynman forces
($F^{\alpha}_{i}$) caused by displacing each ion in each possible
direction ($u^{\beta}_{j}$) in turn.  (Here Greek indices label the
Cartesian coordinates, and $i$ and $j$ run over all the atoms in
the unit cell.) In practice, we take steps $\Delta u$ that are 0.2\% 
in lattice units, 
average over steps in positive and negative directions, and the
resulting $\Phi$ matrix is symmetrized to clean up numerical errors.
The dynamical matrix
$D^{\alpha \beta}_{ij}=(M_iM_j)^{-1/2}\,\Phi^{\alpha \beta}_{ij}$
is then diagonalized to obtain the eigenvalues $\omega^2$.
Once again, we will mainly focus on the monoclinic phase, and briefly
summarize the results for the cubic and tetragonal phases. 

\begin{table}
\caption{ Frequencies (in cm$^{-1}$) of IR-active phonon modes for
ZrO$_{2}$ phases. For monoclinic ZrO$_2$, a possible
reassignment is proposed. Notation `sh' stands for `shoulder' as
in the original reference. Modes labeled `weak' have
very small intensity.  Ref.~\protect\onlinecite{mirgorodsky} is a
previous theoretical work.}
\begin{tabular}{cllll}
\multicolumn{1}{l}{Cubic} & \mbox{This work} & & & \\
\tableline
1 & \mbox{258 ($T_{1u}$)} &      &    &     \\
\tableline
\multicolumn{1}{l}{Tetrag.} & \mbox{This work}
    & \mbox{Expt.~\protect\onlinecite{hirata2}}
    & \mbox{Expt.~\protect\onlinecite{pecharroman}}
    & \mbox{Ref.~\protect\onlinecite{mirgorodsky}} \\
\tableline
1 & \mbox{154 ($E_{u}$)}  & 140 & 164 & 146 \\
2 & \mbox{437 ($E_{u}$)}  & 550 & 467 & 466 \\ 
3 & \mbox{334 ($A_{2u}$)} & 320 & 339 & 274 \\ 
\tableline
\multicolumn{1}{l}{Mono.} & This work 
    & \mbox{Expt.~\protect\onlinecite{feinberg}}
    & \mbox{Expt.~\protect\onlinecite{hirata}}
    & \mbox{Expt.~\protect\onlinecite{zhang99}}\\	 
\tableline
     &                     & \mbox{104} & &\\
  1  & \mbox{181 ($B_{u})^{\rm weak}$} & \mbox{180} & &  \\
     &                     & \mbox{192} & & \\
  2  & \mbox{224 ($A_{u}$)} & \mbox{235} & 220 & 224 \\
  3  & \mbox{242 ($A_{u}$)} &  & &  \\
  4  & \mbox{253 ($B_{u}$)} & \mbox{270} & 250 & 257 \\
  5  & \mbox{305 ($A_{u}$)} &  & &  \\
  6  & \mbox{319 ($B_{u}$)} &            &     & \mbox{324$^{\rm sh}$ (?)} \\
  7  & \mbox{347 ($A_{u}$)} &  &  &  \\
  8  & \mbox{355 ($B_{u}$)} & \mbox{360} & 330 & 351 \\
     &                     & \mbox{375} & 370 & 376 \\
  9  & \mbox{401 ($A_{u}$)} &    &     & \\
  10 & \mbox{414 ($B_{u}$)} & \mbox{415} & 420 & 417 \\
     &                     & \mbox{445} & 440 & \mbox{453$^{\rm sh}$} \\
  11 & \mbox{478 ($A_{u}$)} &  &  &  \\
  12 & \mbox{483 ($B_{u}$)} & \mbox{515} & 520 & 511 \\
  13 & \mbox{571 ($A_{u}$)} & \mbox{620} & 600 & 588 \\
  14 & \mbox{634 ($A_{u})^{\rm weak}$} &            &     & \mbox{687 (?)} \\
     &                     &            &     & \mbox{725 (?)} \\
  15 & \mbox{711 ($B_{u}$)} & \mbox{740} & 740 & 789
\end{tabular}
\label{table:mono_infrared}
\end{table}

The low-temperature phase of ZrO$_{2}$ is monoclinic, with space group
$P2_{1}/c$. The little group at $\Gamma$ is the point group
$C_{2h}$ consisting of operations $E$, $I$, $C_2^y$, and $M_y$.
The character table of this point group indicates that there are
four symmetry classes and thus four irreducible representations,
each of which is one-dimensional.  A standard group-theoretical
analysis indicates that the modes at the $\Gamma$ point can be
decomposed as
\begin{equation}
\label{eqn:mono_modes}
\Gamma^{\rm mono}_{\rm vib} = 9 A_{g} \oplus 9 A_{u} \oplus 9 B_{g} \oplus 9 B_{u}
\end{equation}
(see also Ref.~\onlinecite{asher75}). Of the 36 modes, 18 modes ($9
A_{g} + 9 B_{g}$) are Raman-active and 15 modes ($8 A_{u} + 7 B_{u}$)
are infrared-active, the remaining three modes being the zero-frequency
translational modes.  Only the 15 infrared-active modes contribute to the
lattice dielectric tensor, as discussed in the next subsection.
Similarly, for the tetragonal ZrO$_2$ phase, 
\begin{equation}
\Gamma^{\rm tetra}_{\rm vib} = 1 A_{1g} \oplus 2 A_{2u} \oplus
3 E_{g} \oplus 3 E_{u} \oplus B_{2u} \oplus 2 B_{1g} \;,
\label{eqn:tetra_modes}
\end{equation}
where the $E_u$ and $E_g$ representations are two-dimensional while all
other modes are one-dimensional.  One
$A_{2u}$ mode and one $E_{u}$ pair are acoustic, leaving one
IR-active $A_{2u}$ and two IR-active $E_{u}$ pairs; $A_{1g}$,
$B_{1g}$ and $E_{g}$ are Raman-active, and $B_{2u}$ is silent (see
also Ref.~\onlinecite{bouvier}).  For the cubic phase one finds
\begin{equation}
\Gamma^{\rm cubic}_{\rm vib} = 2 T_{1u} \oplus T_{1g}
\label{eqn:cubic_modes}
\end{equation}
where both $T_{1u}$ and $T_{1g}$ representations are three-dimensional.
One of the $T_{1u}$ triplets is translational, leaving one IR-active
$T_{1u}$ triplet.

\begin{figure}
\begin{center}
   \epsfig{file=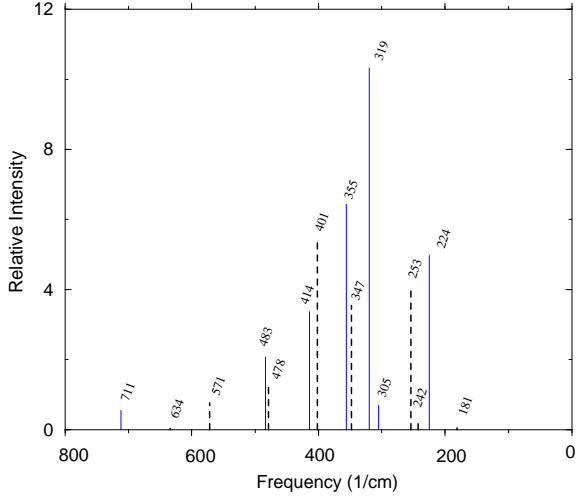, width=3.0in}
\end{center}
\caption{Calculated spectrum of IR-active modes, in which
orientationally-averaged intensity ($\epsilon_{\lambda}$ of
Eq.~(\ref{eq:epsdef})) is plotted vs.~mode frequency in cm$^{-1}$
(see labels on modes).  Solid and dashed lines indicate
$A_{u}$ and $B_{u}$ IR-active modes respectively.}
\label{fig:infrared}
\end{figure}

Table~\ref{table:mono_infrared} lists our calculated IR-active phonon
frequencies in comparison with available theoretical \cite{mirgorodsky} 
and experimental values.\cite{feinberg,hirata,zhang99,hirata2,pecharroman}  
In some cases, possible reassignments are suggested. The overall
agreement is very good; we obtain all the major features of the
experimental infrared spectra. In order to facilitate comparison with
experiment, the oscillator strengths of the infrared-active modes
(namely $\epsilon_{\lambda}$, see Eqs.~(\ref{eq:trace}-\ref{eq:epsdef})
of the Sec.~\ref{sec:dielectrics}) are calculated and plotted
versus frequency in Fig.~\ref{fig:infrared}. The horizontal axis
is reversed for comparison with experimental spectra such as that
of Fig.~2 of Ref.~\onlinecite{hirata}.  The solid and dashed
lines indicate $A_{u}$ and $B_{u}$ modes, respectively.
The two modes at 181\,cm$^{-1}$ and 634\,cm$^{-1}$ are very weak,
so that it is not surprising that they were not observed in most
experiments. The mode at 242\,cm$^{-1}$ is buried by the modes at
253\,cm$^{-1}$ and 224\,cm$^{-1}$, while the mode at 305\,cm$^{-1}$
is similarly shadowed by the strongest mode at 319\,cm$^{-1}$.
Because the pairs of modes at 347/355\,cm$^{-1}$,
401/414\,cm$^{-1}$ and 478/483\,cm$^{-1}$ are very close and of
comparable strength, we think that they might be observed as single
modes in the experiments.

The calculated Raman-active phonon mode frequencies for the
monoclinic structure are summarized in Table~\ref{table:mono_raman}.
The overall pattern of the calculated Raman-active spectrum agrees
quite well with the experimental results, but we again suggest
possible reassignments of some of the modes. Specifically, we
obtained one Raman-active mode at 180\,cm$^{-1}$ that was not observed
in either experiment.  We agree with Carlone \cite{carlone} in excluding
the mode at 355\,cm$^{-1}$ suggested in Ref.~\onlinecite{asher75},
and in interpreting the feature at 780\,cm$^{-1}$ as a first-order
and not a second-order one.\cite{asher75} 
On the other hand, our calculations do not give any frequency
close to 705\,cm$^{-1}$ as observed by Carlone.\cite{carlone} 
The mode at 317\,cm$^{-1}$ obtained in our calculation is observed
somewhat ambiguously in one experiment \cite{asher75} but not in the
other.\cite{carlone} The reason why we assigned the highest
calculated mode at 748\,cm$^{-1}$ as shown in
Table~\ref{table:mono_raman} is that the corresponding
Raman spectra at 15\,K indicated this mode at
745\,cm$^{-1}$.\cite{carlone}  

The overall good correspondence between our results and the experimental
data for both infrared and Raman-active modes therefore tends to
justify our phonon analysis, suggesting that we are now on firm ground
to proceed to the calculation of the lattice
contributions to the dielectric tensors for the ZrO$_{2}$ phases.

\subsection{Lattice Dielectric Tensors}
\label{sec:dielectrics}

In this section, we present our calculations of the lattice
contributions to the static dielectric tensor ($\epsilon_{0}$),
which can be separated into contributions arising from purely
electronic screening ($\epsilon_{\infty}$) and IR-active phonon
modes according to \cite{explan-strain} 
\begin{equation}
\epsilon^{0}_{\alpha \beta} = \epsilon^{\infty}_{\alpha
\beta} + \frac{4 \pi e^{2}}{M_0\,V} \sum_{\lambda} \frac{
 {\widetilde{Z}}^{*}_{\lambda \alpha} \, {\widetilde{Z}}^{*}_{\lambda \beta}
}{\omega_{\lambda}^{2}} \;.
\label{eq:lattcont}
\end{equation}
Here $\alpha$ and $\beta$ label Cartesian coordinates, $e$ is the electron
charge, $M_0$ is a reference mass that we take for convenience to
be 1\,amu, $\omega_{\lambda}$ is the frequency of the $\lambda$-th
IR-active phonon normal mode,
and $V$ is the volume of the three-atom,
six-atom, or 12-atom unit cell for cubic, tetragonal, or monoclinic
cases, respectively.
The mode effective
charge tensors ${\widetilde{Z}}^{*}_{\lambda \alpha}$ are given by
\begin{equation}
\widetilde{Z}^{*}_{\lambda \alpha} = \sum_{i \beta} \, Z^{*}_{i, \alpha
\beta} \, \left(\frac{M_0}{M_i}\right)^{1/2} \, \xi_{i,\lambda \beta}
\end{equation}
where $\xi_{i,\lambda \beta}$, the eigendisplacement of atom $i$ in
phonon mode $\lambda$, is normalized
according to $\sum_{i \alpha} \xi_{i,\lambda \alpha} \,
\xi_{i,\lambda^{'} \alpha} = \delta_{\lambda \lambda^{'}}$.
It is also convenient to write
\begin{equation}
{\rm Tr}[\epsilon^{0}] = {\rm Tr}[\epsilon^{\infty}] + \sum_{\lambda} \epsilon_{\lambda}
\label{eq:trace}
\end{equation}
where
\begin{equation}
 \epsilon_{\lambda}= {4\pi e^2\over M_0 V\omega_{\lambda}^2} \,
 {\widetilde{Z}}_{\lambda}^{*\,2} 
\label{eq:epsdef}
\end{equation}
is the contribution to the trace of the dielectric tensor coming
from the mode $\lambda$, and the scalar mode effective charge
${\widetilde{Z}}^{*}_{\lambda}$ is defined via
${\widetilde{Z}}^{*\,2}_{\lambda} = \sum_\alpha
{\widetilde{Z}}^{*\,2}_{\lambda \alpha}$.

Presented in Table~\ref{table:dielectric} are the scalar mode effective
charges $\widetilde{Z}^{*}_{\lambda}$ and the corresponding contribution to
the static dielectric response $\epsilon_{\lambda}$ for each IR-active mode.
(Note that $T_{1u}$ and $E_u$ modes are three-fold and two-fold
degenerate, respectively. The $\epsilon_{\lambda}$ vs.~$\omega_\lambda$
for the monoclinic phase are also presented graphically in
Fig.~\ref{fig:infrared}.)
From Table~\ref{table:dielectric} or Fig.~\ref{fig:infrared},
we find that for the monoclinic phase the softest modes have small
$\widetilde{Z}^{*}_\lambda$ values and hence do not contribute much
intensity, while the modes with largest $\widetilde{Z}^{*}_\lambda$
are at significantly higher frequency ($\sim$ 319\,cm$^{-1}$).
This observation will be important for explaining the relative
smallness of the dielectric tensor of the monoclinic phase, as discussed
below.

\begin{table}
\caption{Frequencies (cm$^{-1}$) of Raman-active phonon modes
($A_{g}$ and $B_{g}$) in monoclinic ZrO$_{2}$. Experimental
data are measured at 300\,K. The assignment connecting the two sets
of experimental results is adopted from Ref.~\protect\onlinecite{carlone}.
We also adopt the notations introduced by the authors of
Ref.~\protect\onlinecite{asher75}: `ambig'
for `observed ambiguously,' `tetra' for `tetragonal phase,'
`sugg' for `unobserved suggested,' and `2nd' for `second order.'}
\begin{tabular}{cccccl}
  \mbox{Mode} & \mbox{This Work}
& \mbox{Mode} & \mbox{Expt.~\protect\onlinecite{carlone}}
& \mbox{Mode} & \mbox{Expt.~\protect\onlinecite{asher75}} \\
\tableline
  &                     &    &     & 1 & \mbox{92$^{~\rm ambig}$}   \\
1 & \mbox{103 ($A_{g}$)} & 1  &  99 & 2 & 101  \\
  &                     &    &     &   & \mbox{148$^{~\rm tetra}$}  \\
2 & \mbox{175 ($B_{g}$)} & 2  & 177 & 3 & 177  \\
3 & \mbox{180 ($A_{g}$)} &    &     &   &      \\
4 & \mbox{190 ($A_{g}$)} & 3  & 189 & 4 & 189  \\
5 & \mbox{224 ($B_{g}$)} & 4  & 222 & 5 & 222  \\
  &                     & 5  & 270 &   & \mbox{266$^{~\rm tetra}$}  \\
6 & \mbox{313 ($B_{g}$)} & 6  & 305 & 6 & 306  \\
\tableline
7 & \mbox{317 ($A_{g}$)} &    &      & 7   & \mbox{315$^{~\rm ambig}$}  \\
8 & \mbox{330 ($B_{g}$)} & 7  & 331  & 8   & 335  \\
9 & \mbox{345 ($A_{g}$)} & 8  & 343  & 9   & 347  \\
  &                     &    &      & 10  & \mbox{355$^{~\rm sugg}$} \\
10 & \mbox{381 ($A_{g}$)} & 9 & 376  &     &      \\
11 & \mbox{382 ($B_{g}$)} & 10 & 376 & 11  & 382  \\
12 & \mbox{466 ($A_{g}$)} & 11 & 473 & 12  & 476  \\
\tableline
13 & \mbox{489 ($B_{g}$)} & 12 & 498 & 13 & 502 \\
14 & \mbox{533 ($B_{g}$)} & 13 & 534 & 14 & 537 \\
15 & \mbox{548 ($A_{g}$)} & 14 & 557 & 15 & 559 \\
16 & \mbox{601 ($B_{g}$)} & 15 & 613 & 16 & 616 \\
17 & \mbox{631 ($A_{g}$)} & 16 & 633 & 17 & 637 \\
   &                     & 17 & 705 &    &     \\
18 & \mbox{748 ($B_{g}$)} & 18 & 780 &    & \mbox{764$^{~\rm 2nd}$}
\end{tabular}
\label{table:mono_raman}
\end{table}

\begin{table}
\caption{Mode frequency, scalar mode effective charge, and contribution
to the trace of the dielectric tensor for each IR-active mode.}
\begin{tabular}{lldd}
& \mbox{Mode (cm$^{-1}$)} & \mbox{$\widetilde{Z}^{*}_{\lambda}$} & $\epsilon_{\lambda}$ \\
\tableline
\mbox{Cubic} 
& \mbox{258 ($T_{1u}$)} & 1.17 & 31.80 \\
\tableline
\mbox{Tetragonal} 
& \mbox{154 ($E_{u}$)} & 1.03 & 34.29 \\
& \mbox{334 ($A_{2u}$)}& 1.48 & 14.92 \\
& \mbox{437 ($E_{u}$)} & 1.35 &  7.27 \\
\tableline
\mbox{Monoclinic}
& \mbox{181 ($A_{u}$)} & 0.07 & 0.05 \\
& \mbox{224 ($B_{u}$)} & 0.84 & 4.97 \\
& \mbox{242 ($A_{u}$)} & 0.22 & 0.31 \\
& \mbox{253 ($A_{u}$)} & 0.86 & 4.10 \\
& \mbox{305 ($B_{u}$)} & 0.42 & 0.69 \\
& \mbox{319 ($B_{u}$)} & 1.72 & 10.33 \\
& \mbox{347 ($A_{u}$)} & 1.09 & 3.54 \\
& \mbox{355 ($B_{u}$)} & 1.51 & 6.43 \\
& \mbox{401 ($A_{u}$)} & 1.57 & 5.44 \\
& \mbox{414 ($B_{u}$)} & 1.27 & 3.37 \\
& \mbox{478 ($A_{u}$)} & 0.93 & 1.34 \\
& \mbox{483 ($B_{u}$)} & 1.16 & 2.07 \\
& \mbox{571 ($A_{u}$)} & 0.84 & 0.77 \\
& \mbox{634 ($A_{u}$)} & 0.06 & 0.00 \\
& \mbox{711 ($B_{u}$)} & 0.88 & 0.55
\end{tabular} 
\label{table:dielectric}
\end{table}

When all the modes are summed over, we obtain the total lattice
contribution to the static dielectric response (the second term of
Eq.~(\ref{eq:lattcont})).  We find
\[\epsilon^{\rm latt}_{\rm cubic} =  \left( \begin{array}{ccc}
        31.8 & 0 & 0 \\
         0     & 31.8 & 0 \\
         0     &   0    & 31.8
          \end{array} \right) \]

\[\epsilon^{\rm latt}_{\rm tetra} =  \left( \begin{array}{ccc}
        41.6 & 0  & 0 \\
        0  & 41.6 & 0 \\
        0  & 0 & 14.9 \\
          \end{array} \right)  \]

\[\epsilon^{\rm latt}_{\rm mono}  =  \left( \begin{array}{ccc}
        16.7 & 0 & 0.98\\
        0 & 15.6 & 0 \\
        0.98 & 0 & 11.7 \\
          \end{array} \right)  \]
The calculated dielectric tensors have the correct forms expected from
the crystal point group: the cubic one is diagonal and isotropic, the
tetragonal one is diagonal with $\epsilon_{xx}=\epsilon_{yy}
\ne\epsilon_{zz}$, and the monoclinic one is only block-diagonal in
$y$ and $xz$ subspaces.
Our values are also in very good agreement with previous
theoretical calculations for the cubic and tetragonal phases.
Ref.~\onlinecite{rignanese2} reports that $\epsilon^{\rm latt} = 29.77$
for the cubic phase, within about 6\% of our result.
Ref.~\onlinecite{rignanese2} also gives the two independent components of
$\epsilon^{\rm latt}$ in the tetragonal phase as 42.36 and 15.03,
again in excellent agreement with our results, and showing the same
enormous anisotropy.

To compare with experiment, we note that
$\epsilon_{\infty}$ can be estimated from the index of refraction 
$n$, which has been reported experimentally to be about 
2.16 ($n^{2}=\epsilon_{\infty}=4.67$),\cite{nassau} 
2.192 ($\epsilon_{\infty}=4.805$),\cite{french} and 
2.19 ($\epsilon_{\infty}=4.80$) \cite{feinberg} 
for the cubic, tetragonal and monoclinic ZrO$_{2}$ phases, respectively. 
Theoretical works have reported that the orientational average 
$\bar{\epsilon}_{\infty} = 5.75$    
for cubic ZrO$_{2}$,\cite{gonze} and $\epsilon_{\infty}^{\parallel}=5.28$ and 
$\epsilon_{\infty}^{\perp}=5.74$ ($\bar{\epsilon}_{\infty} =
5.59$) for tetragonal ZrO$_{2}$.\cite{rignanese2} 
We can see that $\epsilon_{\infty}$ does not vary strongly
with structural phase, nor is there any evidence for strong
anisotropy.  Moreover, the only experimental measurements of
$\epsilon_0$ of which we are aware are on polycrystalline samples,
for which we need to take an orientational average anyway.
Therefore, we somewhat arbitrarily assume an isotropic value
of $\epsilon_{\infty}$=5.0 for the purposes of comparison with
the total dielectric response.  Then we obtain orientationally
averaged static dielectric constants of
36.8, 46.6 and 19.7 for the cubic, tetragonal and monoclinic
ZrO$_{2}$ phases, respectively.

Experimental reports of the value of $\epsilon_{0}$ for monoclinic
ZrO$_{2}$ span a wide range from about 16 to 25;
\cite{feinberg,rignanese} our estimated value of 19.7 falls
comfortably in the middle of this range.  Unfortunately, we are not aware of
any experimental measurements of the static dielectric response in the
cubic or tetragonal phase.
Since these phases exist only at elevated temperatures, comparison
with zero-temperature theory would need to be made with caution
in any case.  However, neither the cubic--tetragonal nor the
tetragonal--monoclinic transition is ferroelectric in character,
so the influence of the thermal fluctuations on $\epsilon_0$
is is probably not drastic.

\subsection{Discussion}
\label{sec:discussion}

As indicated in the Introduction, much current interest in ZrO$_2$
and related oxides is driven by the search for high-$\epsilon_0$
materials for use as the gate dielectric in future-generation
integrated-circuit devices.  While the dielectric constant of
monoclinic ZrO$_2$ is much bigger than that of SiO$_2$, our results
indicate that it is actually rather low compared to the values in
the range 35-50 expected for the tetragonal and cubic phases.  From
this perspective, it appears that monoclinic ZrO$_2$ has a
disappointingly low static dielectric response.

As can be seen from Eq.~(\ref{eq:lattcont}) or (\ref{eq:epsdef}),
the contribution of a given mode to the dielectric response scales
as $\widetilde{Z}^{*\,2}_\lambda/\omega_\lambda^2$, so that
a large $\epsilon_0$ will result if there are modes that have
simultaneously a large $\widetilde{Z}^{*}$ and a small $\omega$.
As can be seen from Table \ref{table:dielectric}, this is not the
case for monoclinic ZrO$_2$.  Instead, we find that the cluster
of modes with the lowest frequencies ($<250\,$cm$^{-1}$) also
have low $\widetilde{Z}^{*}$ values ($<0.5$), while the most
active modes reside at higher frequencies ($\sim\,300-500\,$cm$^{-1}$).
This is in direct contrast to the case of the cubic perovskite
CaTiO$_3$ studied recently by Cockayne and Burton,\cite{cockayne} 
who find a very soft mode $\omega\simeq100\,$cm$^{-1}$) and very
active ($\widetilde{Z}^{*}\simeq 3$ mode, contributing to an
enormous dielectric constant $\epsilon_0>250$.

The much larger values of $\epsilon_0$ obtained for the
cubic and tetragonal phases suggests that the unfavorable
coincidence of low-$\omega$ and low-$\widetilde{Z}^{*}$ values
may be peculiar to the monoclinic phase, and that other structural
modifications (e.g., quasi-amorphous structures) may actually
have a significantly larger $\epsilon_0$.  This clearly presents
an avenue for future study.

Finally, in low-symmetry structures such as the monoclinic (or
especially amorphous) phases, it is of interest to attempt to
decompose $\epsilon_0$ spatially into contributions coming from
different atoms in the structure.  For example, one might ask
whether it is primarily the three-fold or the four-fold oxygens
that are responsible for the dielectric response in the monoclinic
phase.  For this purpose, we first carry out a decomposition
$\epsilon_{\alpha\beta}^{\rm latt}=\sum_{ij} 
\tilde{\epsilon}^{\,ij}_{\alpha \beta}$ of the
lattice dielectric tensor into contributions
\begin{eqnarray*}
\tilde{\epsilon}^{\,ij}_{\alpha \beta} =
\frac{4 \pi e^{2}}{V} \sum_{\lambda}
\frac{1}{\kappa_{\lambda}}R^{\lambda}_{\alpha i} R^{\lambda}_{\beta j}
\end{eqnarray*}
arising from pairs of atoms,
where $\kappa_{\lambda}$ and $e^{\lambda}_{j \beta}$ are the eigenvalue
and eigenvector of the force constant matrix $\Phi_{ij}^{\alpha \beta}$
for the phonon mode $\lambda$, $V$ is the volume of unit cell, and
$R^{\lambda}_{\alpha j} = \sum_{\beta} Z^{*}_{j, \alpha \beta}\,
e^{\lambda}_{j \beta}$.  We then heuristically define the contribution
coming from atom $i$ to be 
\begin{equation}
\bar{\epsilon}^{\,(i)}_{\alpha \beta}=\sum_{j} \frac{1}{2} 
\left( \tilde{\epsilon}^{\,ij}_{\alpha \beta} +
\tilde{\epsilon}^{\,ji}_{\alpha \beta} \right) \;.
\end{equation}
This atom-by-atom decomposition attributes most of the contribution
to $\epsilon_0$ as coming from the Zr atoms (exactly 2/3 in the
cubic phase and close to this ratio in the other two phases).
As for the oxygen, we found that both the three-fold and four-fold
oxygen atoms make a similar contribution to the orientationally
averaged dielectric constant in the monoclinic phase.  (Not
surprisingly, the anisotropies of the two oxygen contributions are
somewhat different.)  While this analysis has not proven especially
fruitful here, it may be useful in future studies of low-symmetry
(e.g., amorphous) phases.

\section{Conclusion}
\label{sec:conclusion}

In summary, we have investigated here the Born effective charge
tensors, lattice dynamics, and the contributions of the lattice
modes to the dielectric properties of the three ZrO$_{2}$ phases.
The structural parameters, including all internal degrees of freedom
of the three ZrO$_{2}$ phases, are relaxed, and excellent agreement
is achieved with experimental structural refinements and with
previous {\em ab initio} calculations. The observed relative
stability of the ZrO$_{2}$ phases is reproduced in our calculation. The
calculated Born effective charge tensors show anomalously large
values of $Z^{*}$, reflecting a strong dynamic charge transfer as
the bond length varies and indicating a partially covalent nature
of the Zr$-$O bonds. The calculated zone-center phonon mode
frequencies are in good agreement with infrared and Raman
experiments.

Finally, the lattice contributions to the dielectric tensors have
been obtained.  We find that the cubic and tetragonal phases have a
much larger static dielectric response than the monoclinic phase,
with an especially strong anisotropy in the tetragonal structure.
The relatively low $\epsilon_0$ in monoclinic ZrO$_2$ arises
because the few lowest-frequency IR-active modes happen to have
rather small oscillator strengths, while the modes with the
strongest dynamical mode effective charges occur at higher
frequency.  This result, together with the predicted increase of
$\epsilon_0$ in the cubic and tetragonal phases, suggests
that the static dielectric constant is a strong function of the
structural arrangement.  Thus, there may be a prospect for larger
$\epsilon_0$ values in structurally modified (e.g., amorphous)
forms of ZrO$_2$, or in solid solutions of ZrO$_2$ with other
oxides.

\section*{Acknowledgments}
This work was supported by NSF Grant 4-21887. We would like 
to thank E. Garfunkel for useful discussions. One of us (X.Z.) thanks
I. Souza for helpful discussions in connection with the calculation
of Born effective charge tensors.



\begin{references}
\bibitem{explan-gga} We used LDA but not GGA in this work for two
reasons.  First, we wanted to facilitate comparison with previous
DFT studies on ZrO$_2$ that were carried out using LDA.
Second, Ref.~\onlinecite{jomard99} has tested the effect of LDA and 
GGA's (PW91 and PB) on ZrO$_2$.  These authors find that while
the lattice parameters increase slightly with the use of the GGA
($a$ is augmented less than 1\% for the monoclinic ZrO$_2$),
the $b/a$ and $c/a$ ratios and internal structural parameters are almost 
unaffected.
\bibitem{explan-strain} All three ZrO$_2$ phases are non-piezoelectric,
so the fixed-strain and free-stress dielectric tensors are
identical for these systems.
\end{references}
\end{document}